\documentclass[a4paper,fleqn,usenatbib]{mnras}

\usepackage[T1]{fontenc}
\usepackage{ae,aecompl}

\usepackage{graphicx}	
\usepackage{amsmath}	
\usepackage{amssymb}	

\usepackage{newtxtext,newtxmath}

\title[Spectal study of MN112]{Modeling spectra of MN112}

\author[A. Kostenkov et al.]{
    A.~Kostenkov,$^{1, 2}$\thanks{kostenkov@sao.ru}
S.~Fabrika,$^{1, 3}$
O.~Sholukhova,$^{1}$
A.~Sarkisyan,$^{1}$
D.~Bizyaev$^{4, 5}$
\\
$^{1}$Special Astrophysical Observatory, Nizhnij Arkhyz, Russia\\
$^{2}$Saint Petersburg State University, 7/9 Universitetskaya Emb., 199034, Saint Petersburg, Russia\\
$^{3}$Kazan Federal University, Kremlevskaya 18, 420008 Kazan, Russia\\
$^{4}$Apache Point Observatory and New Mexico State University, Sunspot, NM, 88349, USA\\
$^{5}$Sternberg Astronomical Institute, Moscow State University, Moscow, 119992, Russia
\\
}

\date{Accepted XXX. Received YYY; in original form ZZZ}

\pubyear{2020}

\begin{document}
\label{firstpage}
\pagerange{\pageref{firstpage}--\pageref{lastpage}}
\maketitle

\begin{abstract}
MN112 is the Galactic luminous blue variable (LBV) candidate with circumstellar nebula. P Cygni is the first discovered LBV, which was recorded during major eruptions in
the 17th century. The stars have similar spectra with strong emission hydrogen
lines, \ion{He}{i}, \ion{N}{ii}, \ion{Si}{ii}, and \ion{Fe}{iii} lines. 
We present results of the spectroscopic analysis and modeling of MN112 spectra.
We obtained main stellar parameters and chemical abundances of MN112 and compared them with 
those of P Cygni. Atmosphere models were calculated using non-LTE radiative transfer code CMFGEN. 
We have used spectra of MN112 obtained with the 3.5-m telescope at the Observatory of Calar Alto and 3.5-m ARC telescope at the Apache Point Observatory. P Cygni spectra were taken with the 6-m BTA telescope.
We have found the best-fit of the observed spectrum with the model at temperature $T_{\text{eff}}= 15\,200$\,K, clumping-corrected mass-loss rate $\dot{M}f^{-0.5}=5.74 \times 10^{-5}\, M_{\odot}\text{yr}^{-1}$, filling-factor $f=0.4$, luminosity $L=5.77 \times 10^5\, L_{\odot}$ for MN112. The ratio of helium to hydrogen He/H is 0.37 (by the number of atoms) with nitrogen overabundance ($X_\text{N}/ X_{\odot} = 6.8$) and the underabundance of
carbon ($X_\text{C}/ X_{\odot} < 0.1$).
\end{abstract}




\section{Introduction}
Luminous blue variables (LBVs) are stars in a short-term stage of evolution of the massive stars. LBV stars are characterized by a high mass loss rate due to the stellar wind.
The stars show strong spectral and photometric variability on different time scales \citep{Genderen2001AA}.
During the visual maximum, when the temperature decreases to 7000--8000\,K (cold state),\ 
the mass-loss rate increases up to $\sim 10^{-4}\, M_{\odot}\text{yr}^{-1}$ \citep{Humph1994, Stahl2001}.
Spectra of LBV stars at maximum brightness in the visual range
are similar to those of A-F type stars \citep{Massey2007}. When the visual brightness decreases, the temperature of a star can reach more than
35\,000\,K \citep{Clark2005}. In this state the LBV spectra are similar to the spectra of WNLh stars.

Classic LBVs evolved from massive stars with $M>40\,M_{\odot}$, have bolometric values between $-8.5$ 
and $-11.5$ ($\log{L / L_{\odot}} \ga 5.3$). The high mass-loss rate does not allow them to become red supergiants (RSGs).
The less luminous stars with $M_{\text{bol}}$ below $-8.5$ mag have initial masses $\sim 25-40\,M_{\odot}$ and can become RSGs \citep{Humphreys2016}.

MN112 was first discovered by \cite{Gvaramadze2010} and classified as an LBV candidate 
according to the results of studying a ring-like nebula in the IR-range and optical spectroscopy of its central star.
In that paper two possible distances were determined based on the location of MN112 near the open cluster NGC 6823 ($2-3.5$\,kpc) or
in the Perseus arm ($\ga 7$\,kpc).
The first accurate distance estimates $\simeq 6.93^{+2.74}_{-1.81}$ kpc were obtained from the GAIA DR2 survey \citep{Bailer-Jones2018}. 
MN112 does not have S Dor-type variability, its magnitude variations are below 0.2 mag.
In the optical spectrum of MN112 only one forbidden line of [\ion{N}{ii}] $\lambda$5755 is present,
which is probably formed in the outer parts of the wind and can be used as the terminal velocity estimate \citep{Stahl1991}.

P Cygni is the first discovered LBV star and one of the brightest stars
in Galaxy. It was first recorded during major eruptions in 1600 and 1655, when
the visual brightness of P Cygni increases to $3$ mag \citep{Humph1994}.
By the beginning of the 18th century the brightness of P Cygni decreased to 5 magnitude.
Since the first observations in 1715 the visual brightness of P Cygni has been slowly increasing by $0.15 \pm 0.02$ per century at constant luminosity \citep{LamersGroot1992}.
Today P\;Cygni has a brightness of $4.8$ mag, irregularly variable by $\pm (0.02 - 0.03)$ mag on time scales about several years
\citep{Balan2010}. 

The spectroscopic studies of P Cygni were presented in the works \citet{Najarro1997} and \citet{Najarro2001}, where
the luminosity $L=6.1 \times 10^5\, L_{\odot}$, temperature $T_{\text{eff}}=18\,700$\,K,
radius $R=76\, R_{\odot}$ and mass-loss rate $\dot{M}=3.3\times 10^{-5}\, M_{\odot}\text{yr}^{-1}$ were determined.
The first distance estimates of $1.7$ and $1.8$ kpc were obtained from the membership of P Cygni
in the cluster IC 4996, which is a part of the association Cyg OB1 \citep{Lamers1983, Najarro1997}.
The GAIA DR2 distance to P Cygni $\simeq 1.37^{+0.56}_{-0.31}$ kpc is 30\% less 
\citep{Smith2019, Bailer-Jones2018}.
The terminal velocity $V_{\infty}=185\, \text{km\;s}^{-1}$ was measured by many forbidden lines of [\ion{Fe}{ii}], 
the line of [\ion{N}{ii}] $\lambda$5755 \citep{Stahl1991} and [\ion{Ne}{ii}] 12.81 $\mu$, [\ion{Ne}{iii}] 15.55 $\mu$ \citep{Lamers1996}.

In this paper we focused on determining the main stellar parameters and chemical abundances of MN112 and comparing them with those of P Cygni.
It is assumed that MN112, like P Cygni, is the LBV star in the dormant state.
The observed spectra of MN112 were compared with non-local thermodynamic equilibrium (non-LTE) model spectra.
There are several non-LTE radiation transfer codes for modeling extended star atmospheres. For our analysis of the optical spectra we have 
used the iterative non-LTE line-blanketing code CMFGEN \citep{Hillier1998}.
CMFGEN code is actively used to study the spectra of LBV stars (e.g. \citealt{Groh2009b, Mehner2017, Maryeva2018}).

\section{Observations and data reduction}
In this work we have used spectra of MN112 obtained with the Cassegrain Twin Spectrograph (TWIN) of the 3.5-m telescope at the Observatory of Calar Alto (Spain)
on 2009 May 5 in the spectral ranges 3500--5600\,\AA{} and 5300--7600\,\AA{}. The spectral resolution was 3.4\,\AA{} with seeing $\simeq$1.3 arcsec. The near-infrared (NIR) spectra of MN112 are obtained with the TripleSpec \citep{Wilson2004}
spectrograph on the 3.5-m ARC telescope at the Apache Point Observatory (APO, New Mexico, USA) in November of 2012 in the spectral range 0.95--2.46\,\micron{} with mean spectral
resolution 5\,\AA.

In addition, we used spectra of MN112 obtained with the 6-m BTA telescope using the SCORPIO spectrograph
on 2009 June 21, with spectral ranges 4030--5830\,\AA{} and 5730--7500\,\AA{}. The spectral
resolution was 5\,\AA{} and seeing $\simeq$1.5 arcsec.
Data reduction was performed using the standard procedures.

\section{Methods}

CMFGEN code solves the radiative transfer equation in the comoving frame for spherical geometry in conjunction with the statistical equilibrium equations 
and the radiative equilibrium equation for the expanding atmospheres of WR, LBV and O-stars \citep{Hillier1990}. 
The velocity law  \citep{Hillier1989} is characterized by an isothermal effective scale height $h$ in the inner atmosphere and becomes a $\beta$ law in the
wind \citep{Lamers1996}:
\begin{equation}\label{betalaw}
    V(r)=\frac{V_0 + (V_{\infty} - V_0)(1 - R_* / r)^\beta}{1 + (V_0/V_*)\exp[{(R_* - r)/h}]} 
\end{equation}
where $R_*$ is the star radius with Rosseland optical depth $\tau \gtrsim 20$,  $V_{\infty}$ is the terminal velocity, $V_*$ is the velocity at $R_*$, $V_0$ defines velocity where transition between hydrostatic structure and wind occurs.
The code uses a simple filling-factor approach to include clumping in models \citep{Hillier1999}:
\begin{equation}
    f(r) = f_{0} + (1 - f_{0})\exp{\left(-\frac{V(r)}{V_{\text{cl}}}\right)}
\end{equation}

where $f_{0}$ is a filling-factor at a distance corresponding to wind velocity $V_{\text{cl}}$. The volume filling-factor is the ratio of the volume filled
with clumps to the total wind volume. It can be defined as the ratio of the average wind density to the density inside clumps.
The emissivities and opacities scale as the square of the density for thermal processes, while electron scattering scale linearly.
This leads to a reduction of the relative strength of the electron-scattering wings in the strong emission lines, compared to a smooth-wind model \citep{Hillier1991}.
The wind clumping reduces the empirical mass-loss rate of the homogeneous model by a factor of $1/\sqrt{f}$.

All models are prescribed by the stellar radius $R_*$, the stellar luminosity $L_*$, the velocity field $V(r)$, the mass-loss rate $\dot{M}$, the volume filling-factor $f$, and the 
abundances $X_i$ of the included elements. Only the terminal velocity can be independently estimated from the spectrum, for other parameters
it is necessary to construct the corresponding model grids depending on the properties of the object. 

The terminal velocity from the optical spectra of P\,Cygni-like stars can be determined using FWHM measurements of forbidden lines which are formed in outer wind regions with near constant expansion velocity, for example, some [Fe\,II] lines and [N\,II] $\lambda$5755 line \citep{Stahl1991}.

Moreover, the wind velocity estimates can be obtained by a blue shift of absorption component of strong hydrogen and helium emission lines. These estimates are significantly affected by spectral resolution. Moreover, velocity values obtained with this method strongly depend on $\beta$ in the wind velocity law. At high $\beta$ the absorption component of P Cyg profile becomes closer to emission component, while emission component decreases in line width and enhances in intensity at line center \citep{Najarro1997}. In this study we used a simple $\beta$ law (\ref{betalaw}) to describe the structure of the wind.

Independent estimates of MN112 temperature from optical spectra can be determined using ratios of lines of different ionization stages of the same chemical element (e.g. Si\,II/Si\,III). We have used additional criteria for more accurate estimates.

A slight decrease in temperature can lead to a significant change in the ionization structure of the wind and an increase of the opacity
in the resonance Lyman series lines \citep{Najarro1997}. 
As a result, the absorption component of the P Cyg profiles of the Balmer series lines
show a notable dependence on the ionization structure of the wind and can be used
for temperature and mass-loss rate estimates \citep{Groh2011}.

Estimates of the lower temperature limit were based on the presence  of the strong N\,II $\lambda$5001, $\lambda$5047 and $\lambda$6486, $\lambda$6610 lines and temperature-sensitive groups of N\,II $\lambda$5915--5960 and Fe\,III $\lambda$5919--5964 lines in the MN112 spectrum.

We have used Si\,III $\lambda$4554/Si\,II $\lambda$6347 lines ratio, absorption component of H$\beta$ line and temperature-sensitive lines presented above to determine wind ionization structure of MN112.

An increase of the photospheric velocity $V_{0}$ leads to a decrease of the photosphere radius $R_{2/3}$ (radius at Rosseland optical depth $\tau = 2/3$)
and an increase of the effective temperature $T_{\text{eff}}$. In  addition, photospheric radius $R_{2/3}$ (and hence an effective temperature $T_{\text{eff}}$) strongly affected by choice of $\beta$ in high density winds with $V_{\text{phot}}>V_{\text{sonic}}$. An accurate estimates of the photospheric velocity cannot be obtained from the optical spectrum. The photospheric velocity was determined from the modeling of the near infrared hydrogen and \ion{He}{i} lines. The equivalent width of lines in NIR spectral region primary controlled by variations of the continuum flux \citep{Najarro1997b}.

The wind clumping were determined in the optical range by the strength of electron-scattering wings of H$\alpha$, H$\beta$. Clumped mass-loss rate $\dot{M}f^{-0.5}$ was considered constant to keep equivalent widths of hydrogen and helium lines equal in models with different clumping factor.
We have assumed that clumping is dumped at low velocities and radiation instabilities become important at $V_{\text{cl}}=100\, \text{km\;s}^{-1}$ \citep{Owocki1991, Hillier1999}.
The choice of velocity $V_{\text{cl}}$, where clumping should be switched on, is crucial for determining ionization structure at low velocities. For example, decrease in $V_{\text{cl}}$ from 100$\, \text{km\;s}^{-1}$ to 30$\, \text{km\;s}^{-1}$ in our models significantly enhance Fe\,II 5169, Si\,II lines in addition to hydrogen and helium lines, while many N\,II and Fe\,III lines become much weaker. Models with $V_{\text{cl}}=30\, \text{km\;s}^{-1}$ require higher effective temperatures to match N\,II $\lambda$5915--5960 and Fe\,III $\lambda$5919--5964, $\lambda$5127, $\lambda$5156 lines. At higher temperatures hydrogen remains ionized even in outer parts of the wind. These models have poor agreement with observations in absorption component of Balmer series lines. The filling-factor $f$ variations with $V_{\text{cl}}=100\, \text{km\;s}^{-1}$ strongly affect only hydrogen, helium, Fe\,II 5169 and Si\,II lines.

The turbulent velocity $V_{\text{turb}}$ can be estimated by the shift of the strong \ion{He}{i} and hydrogen lines to the red part of the spectrum. In lines with high optical depth (e.g. H$\alpha$, H$\beta$), most emitted line photons are reabsorbed in resonance zone. In this case, there are significant chance that photons will experience electron scattering. As a result, redshifted photons cannot be reabsorbed within resonance zone \citep{Hillier1989}. Moreover, strength of absorption components depends on the turbulent velocity.

We have determined luminosity by approximating photometric data with the model spectrum in different filters. Best-fit model can be scaled to the new luminosity, radius and mass-loss rate with constant effective temperature as $L\sim R^{2}$ and $L\sim \dot{M}^{4/3}$.
There are more accurate relations between these parameters for P\,Cygni like stars \citep{Najarro1997}.

After determining the fundamental parameters ($L$, $T$, $\dot{M}$, $V_{\infty}$), we have found H/He ratio and chemical abundances of C, N, Si, Fe. 
The ratios between H$\alpha$, H$\beta$, H$\gamma$ and He\,I lines were used to find hydrogen fraction in the wind.
We have determined carbon and nitrogen abundances by approximation of C\,II $\lambda$7231, $\lambda$7236 and N\,II $\lambda$5047, $\lambda$5001-5010 lines, respectively.

\section{Results}
\begin{table}
    \caption{Main model parameters and chemical abundances for MN112 and P Cygni; (a) the model parameters were taken from \citet{Najarro2001}; (b) $R_{2/3}$ and $T_{\text{eff}}$ are radius and temperature at $\tau=2/3$, $T_{*}$ is the temperature at hydrostatic radius $R_{*}$ ($\tau \gtrsim 20$) (c) Parameters of reproduced CMFGEN model of P Cygni.}
\label{tab1}
\centering
\begin{tabular}{c c c}
    \hline
   & MN112 & P Cygni \\
  \hline
  	$d$\:[kpc] & $6.93$ & $1.8$\\
    $L_{\ast}\: [L_{\odot}]$ & $5.77 \times 10^{5}$ & $6.1 \times 10^{5}$\\ 
    $R_{*}\: [R_{\odot}]^b$ & $49$ & $76$\\
    $R_{2/3}\: [R_{\odot}]^b$ & $110$ & $105^c$\\
    $T_{*}\: [\text{kK}]^b$ & $22.8$ & $18.7$\\
    $T_{\text{eff}}\: [\text{kK}]^b$ & $15.2\pm0.5$ & 15.9$^c$\\
  $\dot{M}f^{-0.5}\: [M_{\odot}\text{yr}^{-1}]$ & $(5.74\pm1.0) \times 10^{-5}$ & $3.3 \times 10^{-5}$ \\
  $V_{\infty}\: [\text{km s}^{-1}]$ & $300\pm10$ & $185$\\
  $V_{\text{0}}\: [\text{km s}^{-1}]$ & $<15$ & $30$\\
  $V_{\text{turb}}\: [\text{km\;s}^{-1}]$ & $25\pm10$ & $15$\\
  $\beta$ & $3.4\pm0.2$ & $2.5$ \\ 
  $f$ & $0.4\pm0.05$ & $0.5$\\ 
  $X_{\text{He}/X_{\text{H}}}$ & $0.37\pm0.05$ & $0.29$\\
  $X_\text{N}/ X_{\odot}$ & $6.8\pm1.5$ & $6.8$ \\
  $X_\text{C}/ X_{\odot}$ & $<0.1$ & $0.3$ \\
  $X_\text{O}/ X_{\odot}$ & - & $0.2$\\
  $X_\text{Si}/ X_{\odot}$ & $1.0\pm0.1$ & $1.1$\\
  $X_\text{Fe}/ X_{\odot}$ & $1.0\pm0.2$ & $1.0$\\
  Ref & This work & (a) \\
 \hline
\end{tabular} \end{table}

\begin{table}
    \caption{Main stellar parameters of MN112 best-fit models with different terminal velocities.}
\label{tab2}
\centering
\begin{tabular}{c c c c}
    \hline
  & Model 1 & Model 2 & Model 3 \\
  \hline
    $L_{\ast}\: [L_{\odot}]$ & $5.77 \times 10^{5}$ & $3.60 \times 10^{5}$ & $3.09 \times 10^{5}$\\ 
    $R_{*}\: [R_{\odot}]$ & 49 & 49 & 49\\
    $R_{2/3}\: [R_{\odot}]$ & 110 & 88 & 88\\
    $T_{*}\: [\text{kK}]$ & 22.8 & 20.2 & 19.5\\
    $T_{\text{eff}}\: [\text{kK}]$ & 15.2 & 15.1 & 14.5\\
  $\dot{M}f^{-0.5}\: [M_{\odot}\text{yr}^{-1}]$ & $5.74 \times 10^{-5}$ & $3.80 \times 10^{-5}$ & $4.50 \times 10^{-5}$\\
  $V_{\infty}\: [\text{km s}^{-1}]$ & 300 & 325 & 388\\
  $V_{\text{0}}\: [\text{km s}^{-1}]$ & 1 & 20 & 19\\
  $V_{\text{turb}}\: [\text{km\;s}^{-1}]$ & 25 & 25 & 25\\
  $\beta$ & 3.4 & 4.0 & 6.5\\ 
  $f$ & 0.4 & 0.35 & 0.1\\ 
  $X_{\text{H}}\:[\%]$ & 40 & 40 & 40\\
  $X_\text{N}/ X_{\odot}$ & 6.8 & 6.8 & 6.8\\
  $X_\text{C}/ X_{\odot}$ & 0.1 & 0.1 & 0.1\\
  $X_\text{O}/ X_{\odot}$ & 0.1 & 0.1 & 0.1\\
  $X_\text{Si}/ X_{\odot}$ & 1.0 & 1.0 & 1.0\\
  $X_\text{Fe}/ X_{\odot}$ & 1.0 & 1.0 & 1.0\\
 \hline
\end{tabular} \end{table}

\begin{figure*}
\centering
    \includegraphics[width=17cm]{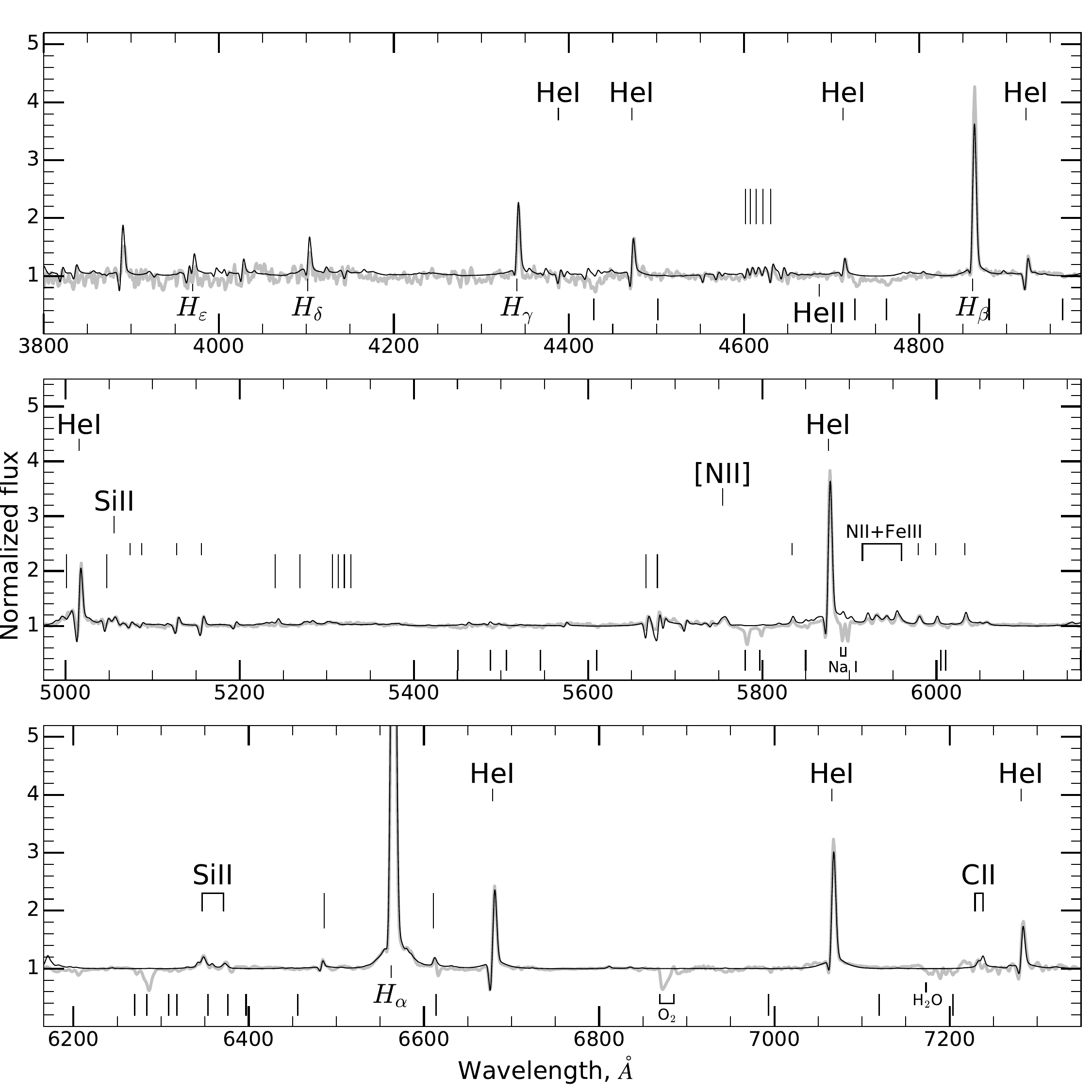}
    \caption{The normalized optical spectrum of MN112 (grey solid line) compared with the best-fit CMFGEN model (black solid line).
        The vertical long lines above the spectrum are \ion{N}{ii}, the short lines are \ion{Fe}{iii}.
        Diffuse interstellar bands (DIBs) under the spectrum are marked with black lines.
    }

    \label{MN112}
\end{figure*}

\begin{figure*}
\centering
    \includegraphics[width=17cm]{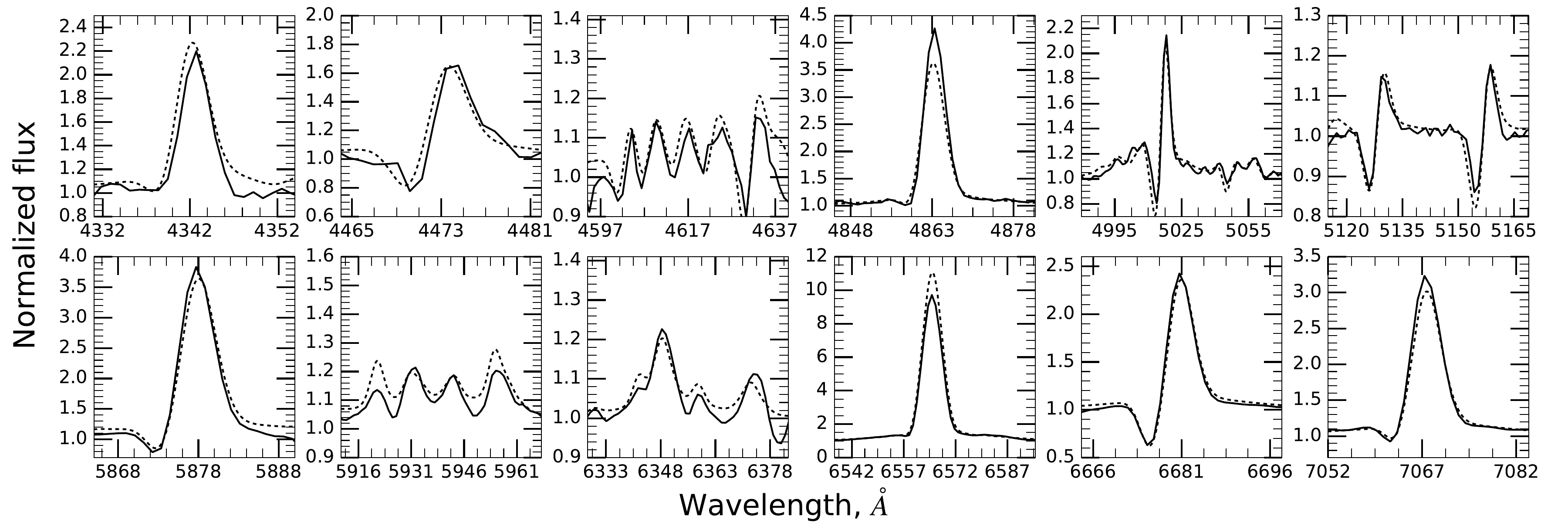}
    \caption{Comparison of selected optical lines in the observed spectrum of MN112 (black sold line) and the best-fit CMFGEN model (black dotted line). Top, from left to right: H$\gamma$; He\,I $\lambda$4471; N\,II $\lambda$4601-4643; H$\beta$; He\,I 5015, N\,II $\lambda$5045, Si\,II $\lambda$5040, $\lambda$5055; Fe\,III $\lambda$5127, $\lambda$5156. Bottom: He\,I $\lambda$5876; \ion{Si}{ii} $\lambda$5916, $\lambda$5959--5961, \ion{N}{ii} $\lambda$5915--5960, \ion{Fe}{iii} $\lambda$5919--5964; Si\,II $\lambda$6348, $\lambda$6371; H$\alpha$; He\,I $\lambda$6678, He\,I $\lambda$7065.}

    \label{MN112_selected_lines}
\end{figure*}

\begin{figure*}
\centering
    \includegraphics[width=17cm]{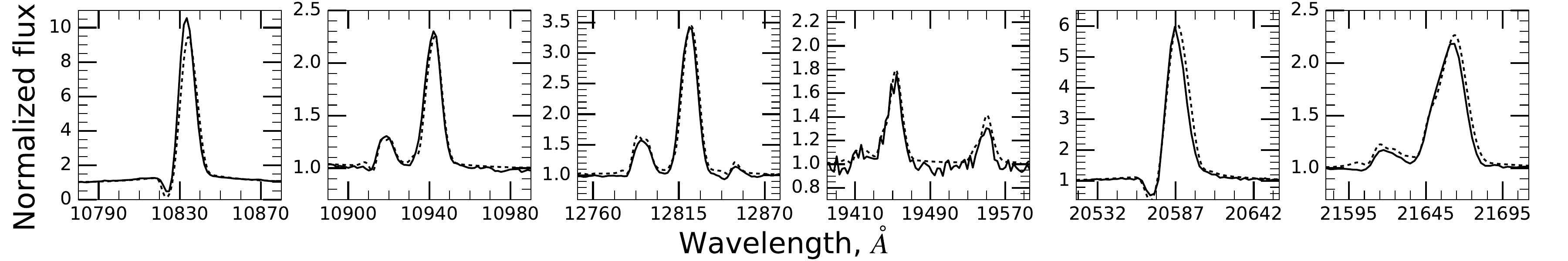}
    \caption{Comparison of selected near-infrared lines in the observed spectrum of MN112 (black solid line) and the best-fit CMFGEN model (black dotted line). From left to right: He\,I $\lambda$10830; He\,I $\lambda$10913, $\lambda$10917, H\,I (Pa$\gamma$) $\lambda$10938; He\,I $\lambda$12783-12846, H\,I (Pa$\beta$) $\lambda$12818; He\,I $\lambda$19434-19543, H\,I (Br$\delta$) $\lambda$19445; He\,I $\lambda$20581; He\,I $\lambda$21617-21649, H\,I (Br$\gamma$) $\lambda$21655.}
    \label{MN112_NIR_selected_lines}
\end{figure*}

\begin{figure}
\centering
\resizebox{\hsize}{!}{\includegraphics{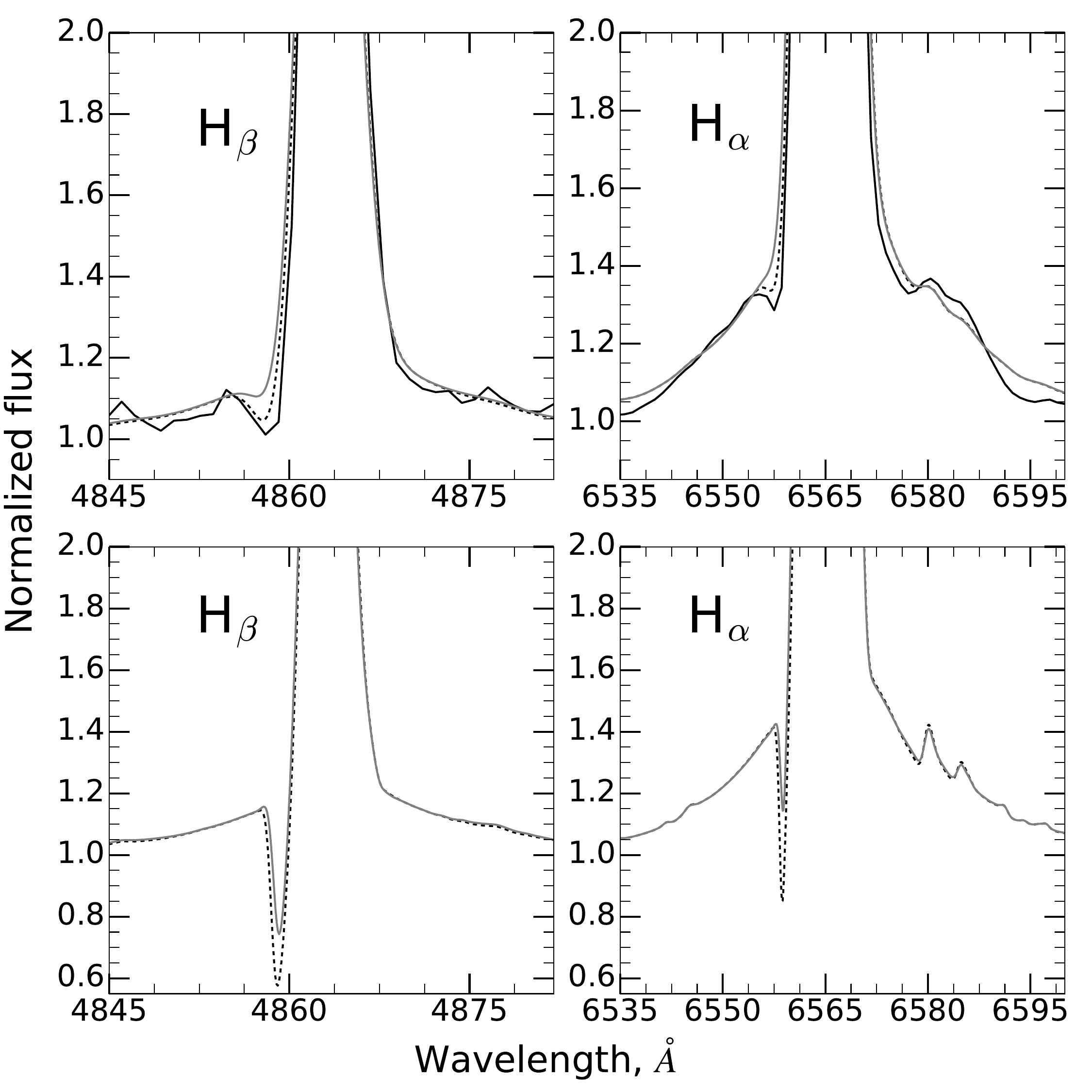}}
     \caption{The absorption profile of H$\alpha$ and H$\beta$ lines in model spectra of MN112 with different temperatures. The observed spectrum of MN112 is marked with black solid line. Best-fit model with effective temperature $T_{\text{eff}}=15.2\,$kK is black dotted line. Model with higher temperature $T_{\text{eff}}=15.5\,$kK is gray solid line Top: model spectra smoothed with spectral resolution that correspond observed spectra (3.4\AA{}). Bottom: model spectra with high spectral resolution 0.1\AA{}.}
      \label{MN112_hyd_abs}
\end{figure}

\begin{figure}
\centering
\resizebox{\hsize}{!}{\includegraphics{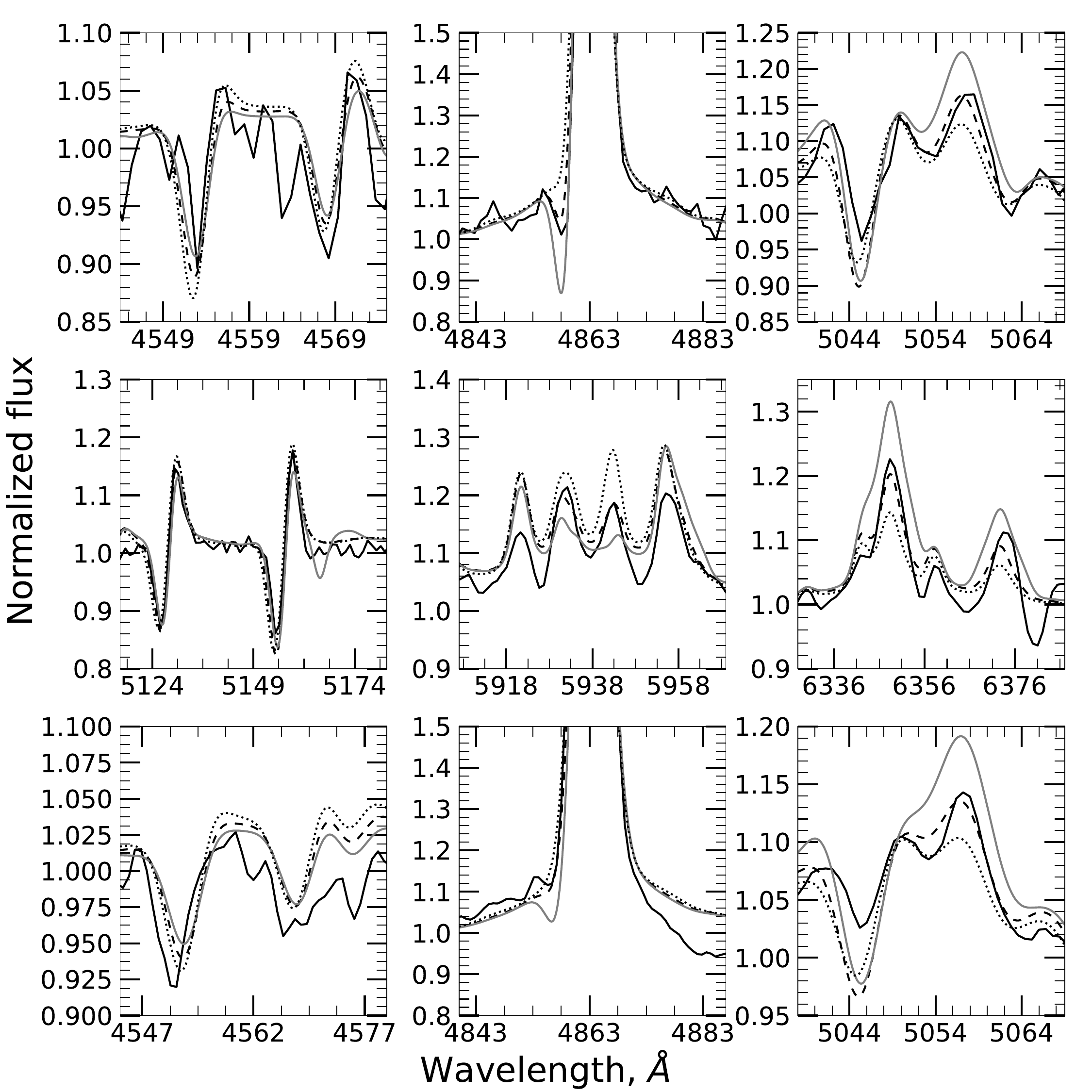}}
     \caption{Model spectra of MN112 smoothed with spectral resolutions 3.4\AA{} (top, mid) and 5\AA{} (bottom) at different temperatures. Observed spectra of MN112 is marked with black solid line. Best-fit model with effective temperature $T_{\text{eff}}=15.2\,$kK is black dashed line. Model with lower temperature $T_{\text{eff}}=14.9\,$kK is grey solid line. Model with higher temperature $T_{\text{eff}}=15.8\,$kK is black dotted line. Top and bottom, from left to right: Si\,III $\lambda$4554, $\lambda$4567; absorption component of H$\beta$ line;  N\,II $\lambda$5045, Si\,II $\lambda$5040. Mid: Fe\,III $\lambda$5127, $\lambda$5156, Fe\,II $\lambda$5169;  \ion{Si}{ii} $\lambda$5916, $\lambda$5959--5961, \ion{N}{ii} $\lambda$5915--5960, \ion{Fe}{iii} $\lambda$5919--5964;  Si\,II $\lambda$6348, $\lambda$6371.}
      \label{MN112_temp}
\end{figure}

\begin{figure*}
\centering
    \includegraphics[width=17cm]{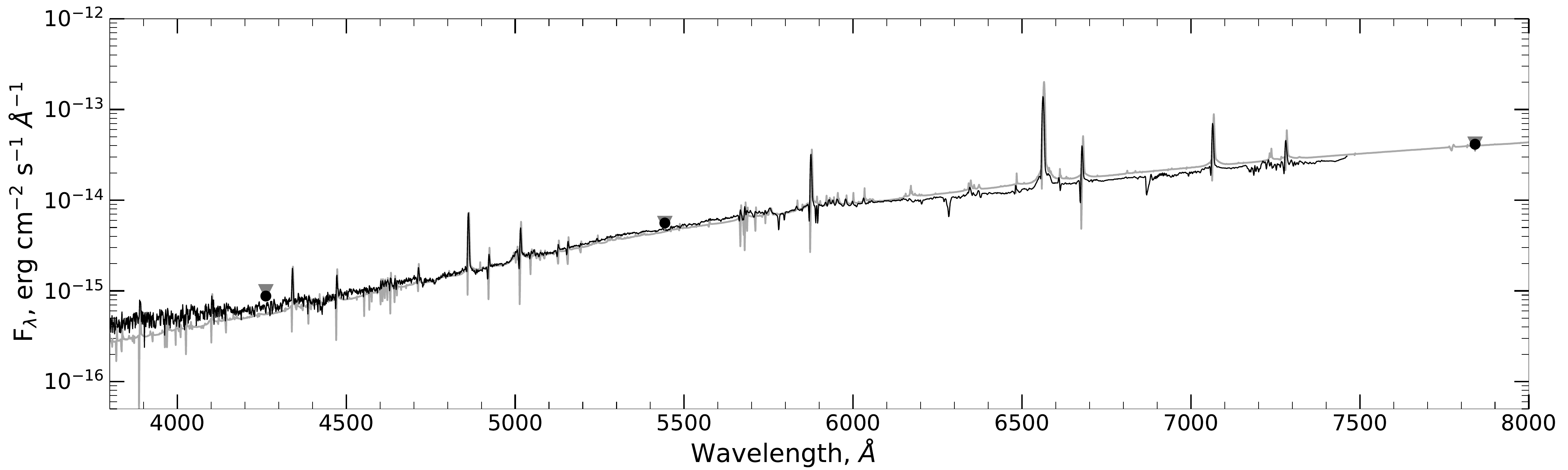}
    \caption{The optical flux-calibrated spectrum of MN112 (black solid line) with best-fit model (grey solid line). Photometric magnitudes in B, V, I bands for our best-fit model are marked with grey triangles. The black circles are photometric magnitudes from \citet{Gvaramadze2010}. Data errors are comparable to the marker size.}
    \label{MN112_fit_phot}
\end{figure*}

\begin{figure*}
\centering
    \includegraphics[width=17cm]{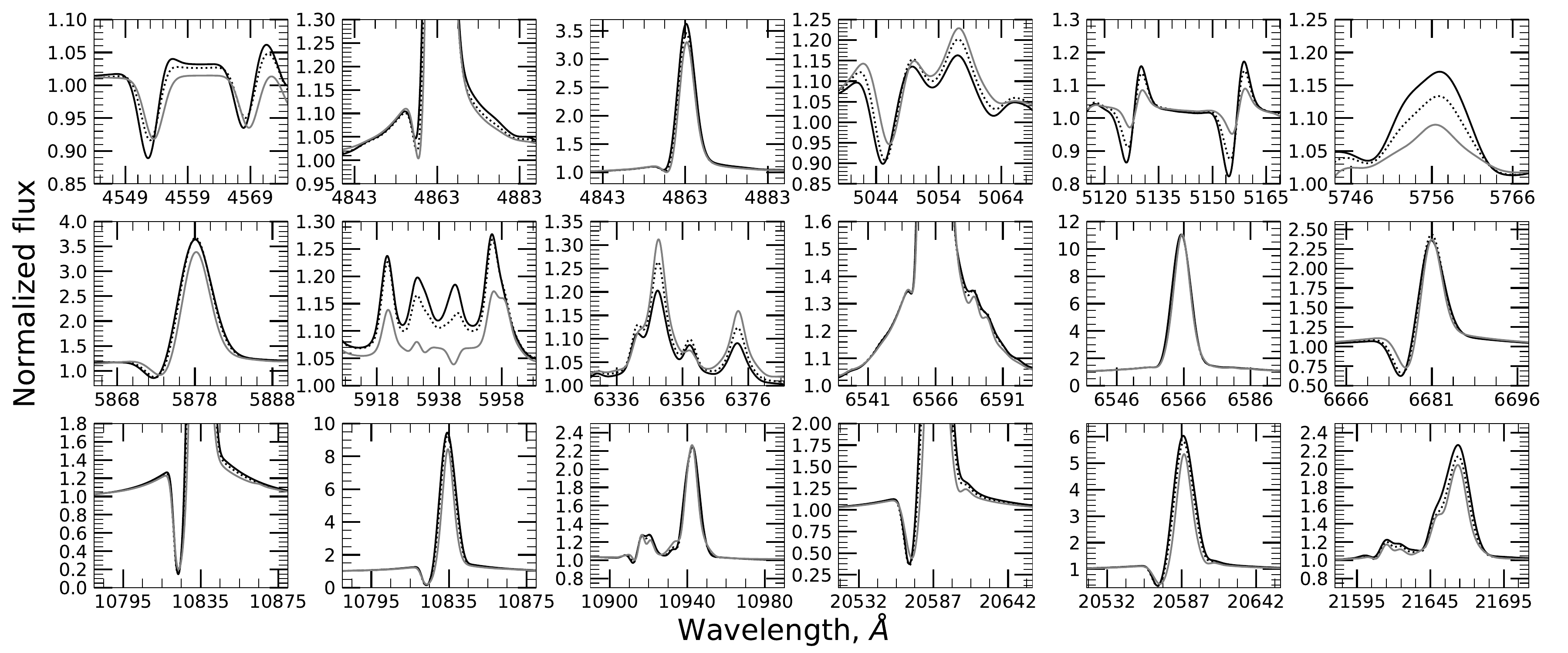}
    \caption{Comparison of selected optical and near-infrared lines in best-fit models with different terminal velocities. Models are marked with black solid ($V_{\infty}=300\, \text{km\;s}^{-1}$), black dotted ($V_{\infty}=325\, \text{km\;s}^{-1}$) and grey solid ($V_{\infty}=388\, \text{km\;s}^{-1}$) lines. Top, from left to right: Si\,III $\lambda$4554, $\lambda$4567; absorption component of H$\beta$ line; H$\beta$; N\,II $\lambda$5045, Si\,II $\lambda$5040;  Fe\,III $\lambda$5127, $\lambda$5156; [N\,II] $\lambda$5755. Mid: He\,I $\lambda$5876; \ion{Si}{ii} $\lambda$5916, $\lambda$5959--5961, \ion{N}{ii} $\lambda$5915--5960, \ion{Fe}{iii} $\lambda$5919--5964; Si\,II $\lambda$6348, $\lambda$6371; absorption component of H$\alpha$ line; H$\alpha$; He\,I $\lambda$6678. Bottom: absorption component of  He\,I $\lambda$10830 line;  He\,I $\lambda$10830; He\,I $\lambda$10913, $\lambda$10917, H\,I (Pa$\gamma$) $\lambda$10938; absorption component of He\,I $\lambda$20581 line; He\,I $\lambda$20581; He\,I $\lambda$21617-21649, H\,I (Br$\gamma$) $\lambda$21655.}

    \label{MN112_diff_models}
\end{figure*}

\begin{figure}
\centering
\resizebox{\hsize}{!}{\includegraphics{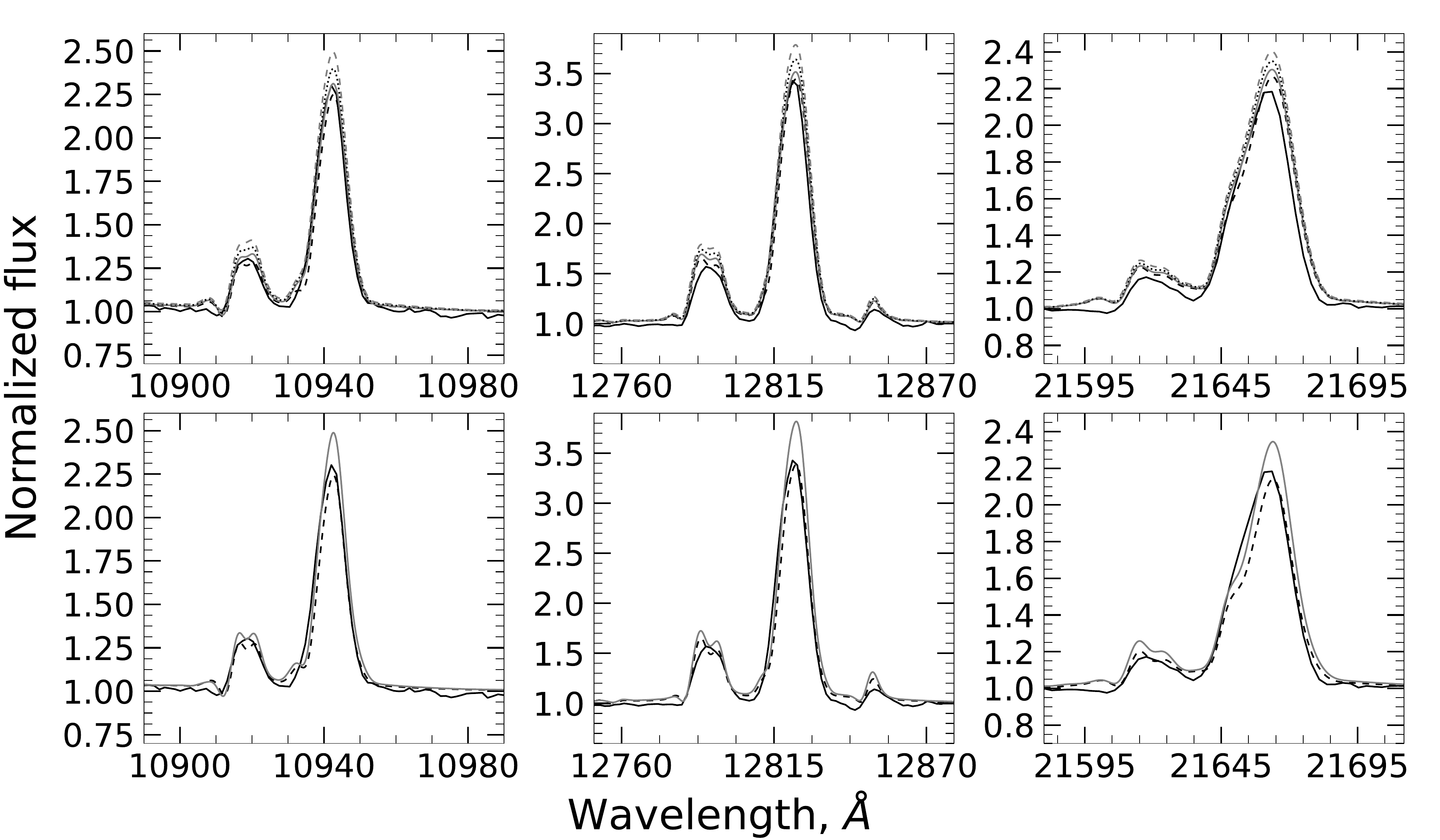}}
     \caption{Best-fit models of MN112 with equal effective temperatures $T_{\text{eff}}=15.2\,$kK at different photospheric velocities. Observed spectrum of MN112 is marked with black solid line. Top: models with terminal velocity $V_{\infty}=300\,\text{km s}^{-1}$; different photospheric velocities are marked with black dashed line ($V_{0}=1\,\text{km s}^{-1}$), grey solid line ($V_{0}=10\,\text{km s}^{-1}$), black dotted line ($V_{0}=15\,\text{km s}^{-1}$) and grey dashed line ($V_{0}=20\,\text{km s}^{-1}$)  Bottom: models with terminal velocity $V_{\infty}=325\,\text{km s}^{-1}$; different photospheric velocities are marked with black dashed line ($V_{0}=20\,\text{km s}^{-1}$) and grey solid line ($V_{0}=30\,\text{km s}^{-1}$). Top and bottom, from left to right: $\lambda$10917, H\,I (Pa$\gamma$) $\lambda$10938;  He\,I $\lambda$12783-12846, H\,I (Pa$\beta$) $\lambda$12818; He\,I $\lambda$21617-21649, H\,I (Br$\gamma$) $\lambda$21655.}
      \label{MN112_vphot}
\end{figure}

\begin{figure}
\centering
\resizebox{\hsize}{!}{\includegraphics{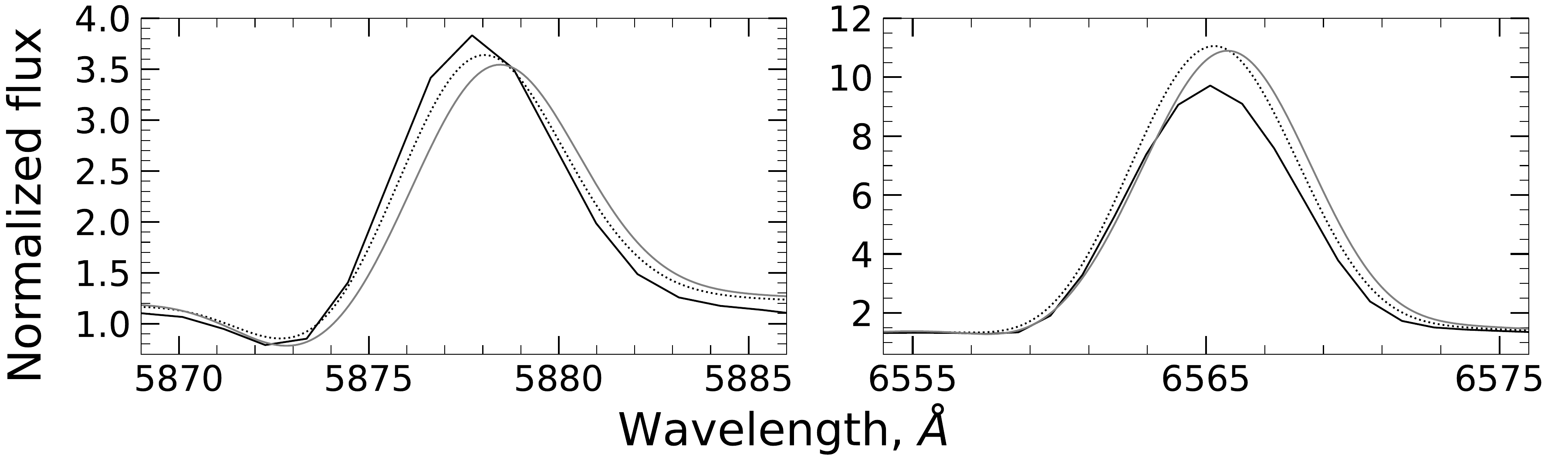}}
     \caption{Comparison of best-fit models ($V_{\infty}=300\,\text{km s}^{-1}$) of MN112 with turbulent velocities $V_{\text{vturb}}=25\,\text{km s}^{-1}$ (black dotted line) and $V_{\text{vturb}}=40\,\text{km s}^{-1}$ (grey solid line). Observed spectrum of MN112 is marked with black solid line.}
      \label{MN112_vturb}
      
\end{figure}

We present main stellar parameters of our best-fit model for MN112 in Table \ref{tab1}.
The parameters from \citet{Najarro2001} were chosen as the best-fit model for P Cygni.
The normalized\footnote{DIBs at red wing of H$\beta$ line were corrected in normalized TWIN spectrum of MN112 for clear comparison with model spectra.} optical spectra and best-fit model of MN112 with the most important lines are illustrated in Fig.\,\ref{MN112}. Selected optical and NIR lines in observed spectra and best-fit model are presented in Fig.\,\ref{MN112_selected_lines} and \ref{MN112_NIR_selected_lines}, accordingly. We have used P Cygni spectrum presented in \citet{Gvaramadze2010} for comparison with spectra of MN112. P Cygni and MN112 have a lot of similar emission lines of hydrogen, \ion{He}{i}, \ion{N}{ii}, \ion{Si}{ii}
and \ion{Fe}{iii} in the optical spectra. Both stars have weak absorption components in the Balmer series lines and strong P Cyg profiles in \ion{He}{i}, \ion{N}{ii} 
and \ion{Fe}{iii} lines. We used several models to investigate how absorption component was resolved in low resolution spectra. Fig.\,\ref{MN112_hyd_abs} shows profile of H$\alpha$ and H$\beta$ lines of our best-fit model with different temperatures at low and high resolutions. Absorption components in hotter model (e.g. small "bump" in H$\alpha$ line) almost disappeared at low resolution and only asymmetric wings of H$\beta$ line indicate to absorption component. Further increase in temperature do not lead to significant changes in H$\alpha$ profile with low resolution.

The spectrum of MN112 has stronger \ion{N}{ii} (e.g. $\lambda$5047, $\lambda$5001-5010, $\lambda$5915--5960, $\lambda$6486, $\lambda$6610), Si\,II ($\lambda$5056, $\lambda$6348, $\lambda$6371)
and \ion{Fe}{iii} ($\lambda$5127, $\lambda$5156, $\lambda$5919--5964) emission lines than P Cygni. There are no \ion{He}{ii} and \ion{N}{iii} emission lines in both spectra. The Oxygen lines are absent in the optical spectra of MN112 and P Cygni. There are only two weak \ion{C}{ii} $\lambda$7231, $\lambda$7236 lines in both spectra. The carbon lines located at the right edge of MN112 and P Cygni spectra. An accurate estimates of carbon abundance cannot be obtained due to low signal-to-noise ratio in this part of the spectra.
There are no \ion{Fe}{ii} and [\ion{Fe}{ii}] in the MN112 spectrum.

For all models we used fixed radius $R=49\,R_{\odot}$. Changes of effective temperature were controlled via luminosity variations. We have used two optical spectra of MN112 for temperature estimates. The spectrum of MN112 obtained with BTA-telescope have lower resolution 5\AA{}, but higher signal-to-noise ratio compared to TWIN spectrum. In Fig\,\ref{MN112_temp} we present best-fit model with different temperatures compared to observed spectra. As seen in Fig\,\ref{MN112_temp}, even small decrease in temperature ($\approx$1\,kK) can significantly increase strength of H$\beta$ absorption component and \ion{Si}{ii}, \ion{Fe}{ii} lines. Lines of \ion{Si}{ii} and \ion{Fe}{ii} strongly depends on the wind density and ionization structure, a large contribution to the abundances of \ion{Si}{ii} and \ion{Fe}{ii} is made by the charge-exchange process in the outer parts of the wind.

Lines of \ion{Fe}{iii} $\lambda$5127, $\lambda$5156 do not show a notable temperature dependence, while the \ion{Fe}{ii} $\lambda$5169 significantly increases at lower temperatures. The groups of \ion{N}{ii} and \ion{Fe}{iii} lines (especially \ion{N}{ii} $\lambda$5915–5960 lines) in the range of $\lambda5915\text{--}5964$ are strongly influenced by variations of the wind ionization structure.

We built more than 100 models in the range 
$\dot{M}=3.0\text{--}7.0\times 10^{-5}\, M_{\odot}\text{yr}^{-1}$ and $T_{*}=19\,000\text{--}23\,000$\,K with different velocity distributions (see below) to more accurately determine temperature and 
mass-loss rate. We found that the temperature of MN112 is
 $T_{\text{eff}}=15200\pm500$\,K and mass-loss rate is $\dot{M}=(5.47\pm1.0) \times
 10^{-5}\,M_{\odot}\text{yr}^{-1}$

The interstellar reddening $A_{\text{v}}=8.45\pm0.11$ was estimated by approximating photometric data in B=$17.13\pm0.12$\,mag, V=$14.53\pm0.03$\,mag, I=$11.15\pm0.03$\,mag bands \citep{Gvaramadze2010} with the best-fit model at $T_{\text{eff}}=15\,200\,$K using \citet{Fitzpatrick1999} absorption curves. The flux-calibrated spectrum of MN112 with best-fit model and photometric magnitudes are presented in Fig.\,\ref{MN112_fit_phot}.

For the distance of $6.93^{+2.74}_{-1.81}$\,kpc, the luminosity of MN112 is $L\simeq 5.77^{+2.28}_{-1.50} \times 10^{5}\, L_{\odot}$ and mass-loss rate is $\dot{M}f^{-0.5}= 5.74^{+1.63}_{-1.17} \times 10^{-5}\,M_{\odot}\text{yr}^{-1}$. In addition, GAIA DR2 distance estimates have errors for stars with large radii related to the light beam shift \citep{Berlanas2019}. For radius $R=42\,R_{\odot}$ ($L\simeq 4.27 \times 10^{5}\, L_{\odot}$, $d=5.12$\,kpc) GAIA DR2 error is $\approx 20\%$. In this case, minimal distance estimate for MN112 according to GAIA DR2 is $\approx 4.0$\,kpc. We suggest that MN112 most probably located at distance $\gtrsim 7$\,pc (in the Perseus arm), because lower distance estimate is higher than upper distance estimate $\approx3.5$\,kpc \citep{Sagar1981} to open cluster NGC~6834 and OB association Vul\,OB1 (in the Orion arm). In addition, more recent researches point to smaller distance values $\approx 2\,$kpc to Vul\,OB1 \citep{Kharchenko2005, Billot2010}. All luminosity estimates presented above depend on the adopted terminal velocity value.

We received FWHM measurements $8.19\pm0.35$\,\AA{} of the [\ion{N}{ii}] $\lambda$5755 line.
It corresponds to the terminal velocity $V_{\infty}=388\pm20\, \text{km\;s}^{-1}$ with correction for instrumental resolution $\simeq3.4$\,\AA{}. These estimates are identical to the wind velocity $V_{\infty}=394\pm50\, \text{km\;s}^{-1}$ obtained in \citet{Gvaramadze2010}. We compared several models with different terminal velocities. Calculations show that models with velocities $300-400\, \text{km\;s}^{-1}$ corrected for instrumental resolution have acceptable agreement with observations in [\ion{N}{ii}] $\lambda$5755 line.

We have calculated three models with different terminal velocities  $V_{\infty}=300\, \text{km\;s}^{-1}$,  $V_{\infty}=325\, \text{km\;s}^{-1}$ and  $V_{\infty}=388\, \text{km\;s}^{-1}$. Main stellar and wind parameters for these models presented in Table\,\ref{tab2}. Fig\,\ref{MN112_diff_models} shows comparison of selected line profiles of different models. We have varied some parameters ($T_{*}$, $\beta$, $f$, $V_{\text{vphot}}$) to obtain equal EW and profile of H$\alpha$ line in all models.

We have investigated the influence of the velocity law on the optical spectrum in our models. Models with lower terminal velocities require lower $\beta$ to match FWHM of H$\alpha$, however, low $\beta$ values significantly affect ionization structure at velocities $V < V_{\infty}/2$. In addition, models with higher $\beta$ have stronger electron scattering wings compared to models with equal emission peak of H$\alpha$ line. It can be probably related to slower drop in density at low velocities. Different filling-factor $f$ values were used to match electron-scattering wings in H$\alpha$ line. We calculated several models for each velocity with a constant empirical mass-loss rate $\dot{M}f^{-0.5}$ to estimate the volume filling-factors. Our filling-factor estimates are typical for LBV stars which ranging from $0.1$ to $0.5$ \citep{Groh2009b, Mahy2016}. We assume that clumping starts at $V_{\text{cl}}=100\, \text{km\;s}^{-1}$, Consequently, variations in filling-factor values affect mass-loss rate ($\dot{M}f^{-0.5}$ should be fixed to keep EW of hydrogen and He\,I lines equal) and hence ionization structure at velocities $V <100\, \text{km\;s}^{-1}$.

Fig\,\ref{MN112_diff_models} shows major difference between models in Fe\,III $\lambda$5127, $\lambda$5156 lines, groups of N\,II $\lambda$5915--5960 and Fe\,III $\lambda$5919--5964 lines. Model with terminal velocity $V_{\infty}=388\, \text{km\;s}^{-1}$ require higher temperature to match N\,II and Fe\,III lines. In this case, mass-loss rate should be increased to keep equal ionization structure in outer parts of the wind and match absorption component of Balmer series lines. As a result, EWs of H$\alpha$ and Si\,II lines become too high and have poor agreement with observations. Model with $V_{\infty}=300\, \text{km\;s}^{-1}$ have best-fit in lines in both inner and outer regions of the wind (Si\,III/Si\,II, Fe\,III and hydrogen lines).

In Fig\,\ref{MN112_vphot} we present selected lines profiles for our best-fit models with different photospheric velocities. The temperature at hydrostatic radius $T_{*}$ was corrected to keep equal ionization structure between models. EW of NIR lines significantly increases at $V_{\text{vphot}} \gtrsim 15\, \text{km\;s}^{-1}$ and $V_{\text{vphot}} \gtrsim 30\, \text{km\;s}^{-1}$ for models with $V_{\infty}=300\, \text{km\;s}^{-1}$ and $V_{\infty}=325\, \text{km\;s}^{-1}$ respectively. Photospheric velocity estimates are $1-15\, \text{km\;s}^{-1}$ for $V_{\infty}=300\, \text{km\;s}^{-1}$ and $20\pm5\, \text{km\;s}^{-1}$ for $V_{\infty}=325\, \text{km\;s}^{-1}$.

The turbulent velocity $V_{\text{turb}}=25\, \text{km\;s}^{-1}$ was determined from the shift of 
the strong \ion{He}{i} and hydrogen lines in models compared to observed spectrum of MN112.
The Balmer series and \ion{He}{i} lines of model spectra were shifted towards long-wavelength region  relative to the observed spectra at turbulent velocities higher than 40$\, \text{km\;s}^{-1}$. Models with different turbulent velocities and observed spectrum are presented in Fig.\,\ref{MN112_vturb}. The turbulent velocity estimates are $V_{\text{turb}}=25\pm10\, \text{km\;s}^{-1}$. In addition, turbulent velocity variations slightly affect absorption component in Si\,III and Balmer series lines, which increases error of temperature estimates.

We found a similarity between the chemical abundances of MN112 and P Cygni. The hydrogen abundance of MN112 is slightly lower: $\text{He/H}\simeq 0.37$\footnote{Wrong value of He/H ratio (0.27 by number of atoms instead of 0.37) was accidentally presented in the published paper.}$\,\pm\:0.05$ (by number of atoms) and $0.29$ for P Cygni, however, both stars have the same nitrogen overabundance ($X_\text{N}/ X_{\odot} = 6.8\pm1.5$) and the underabundance of carbon ($X_\text{C}/ X_{\odot} < 0.1$ for MN112 and $X_\text{C}/ X_{\odot} = 0.3$ for P\,Cygni). We used solar metallicity ($X_\text{Si}/ X_{\odot} = 1.0\pm0.1$, $X_\text{Fe}/ X_{\odot} = 1.0\pm0.2$) for all MN112 models.

\section{Discussion and conclusions}

In this paper, we present the results of modeling with CMFGEN code the optical and NIR spectra of Galactic cLBV star.
According to the modeling results we have achieved good consistency of synthetic spectra with observations. 

The first estimates of the luminosity and basic stellar parameters of MN112 were obtained.
The hydrogen fraction of MN112 by number of atoms is slightly lower ($\text{He/H}\simeq 0.37$) compared to P Cygni ($\text{He/H}\simeq 0.29$),
while the chemical abundances of N, Si and Fe are almost equal.
In the MN112 spectrum the Balmer series lines have weak absorption components and the \ion{Fe}{ii} lines are not present.
We found that MN112 has higher temperature at hydrostatic radius $T_{*}=22\,800\,$K and lower effective temperature $T_{\text{eff}}=15\,200\,$K than P Cygni. Significant difference between $T_{*}$ and $T_{\text{eff}}$ indicates to highly extended photosphere of MN112.

The relative strength of the electron-scattering wings in the Balmer series lines in all models of MN112 (volume filling-factor $f=0.1-0.4$) is less compared to P Cygni ($f=0.5$).
The luminosity of out best-fit model of MN112 $L=5.77 \times 10^5\, L_{\odot}$ close to the luminosity of P Cygni.
We obtain the mass-loss rate  $\dot{M}f^{-0.5}=5.74 \times 10^{-5}\, M_{\odot}\text{yr}^{-1}$, velocity law $\beta=3.4$, 
photospheric velocity $V_{0}=1\, \text{km\;s}^{-1}$ and terminal velocity $V_{\infty}=300\, \text{km\;s}^{-1}$ estimates.
The wind performance number of MN112 $\eta=c \dot{M} V_{\infty} / (L_{*})$ is almost equal to P\,Cygni one ($V_{\infty}=185\, \text{km\;s}^{-1}$). MN112 are in slightly higher ionization state even with lower temperature compared to P\,Cygni, which related to significant difference in filling-factors. In case of determination of mass-loss rate by recombination hydrogen and He\,I lines, increase of filling-factor reduce wind density at low velocities $V<V_{\text{cl}}$.

We have compared the spectrum of MN112 with the spectra of confirmed LBVs, LBVs candidates and Of/late-WN stars presented in \citet{Humphreys2014}. There are three LBVs AF And in M31
and Var B, Var 2 in M33 that have the optical spectrum is similar to MN112. The spectra of these stars do not have \ion{Fe}{ii} and [\ion{Fe}{ii}] lines and they are similar to the spectra of Of/late-WN stars.
As shown by \citet{Humphreys2017}, LBVs have lower wind velocities $V=229\, \text{km\;s}^{-1}$ for hydrogen lines and $V=221\, \text{km\;s}^{-1}$ for He I lines compared to $V=329\, \text{km\;s}^{-1}$ and $V=313\, \text{km\;s}^{-1}$ for Of/late-WN stars. The wind velocities of MN112 are
 $V=263\, \text{km\;s}^{-1}$ for hydrogen and $V=187\, \text{km\;s}^{-1}$ for \ion{He}{i} lines. These estimates close to the wind velocities of LBV stars.
However, MN112 has a higher wind terminal velocity estimates $V_{\infty}=300-400\, \text{km\;s}^{-1}$ with an extended wind acceleration zone ($\beta=3.4-6.5$).

The effective temperature and wind terminal velocity of MN112 are lower than those of the LBV candidates with Of/late-WN like star spectra, while MN112 has a significantly higher mass-loss rate compared to late B-type stars. The presence of short-term spectroscopic changes could confirm the LBV status of MN112. We have compared the spectrum of MN112 from this work with the spectrum obtained in 2015 with 6-m BTA telescope. Unfortunately, low spectral resolution ($\sim 5$\,\AA{}) does not allow us to estimate the changes in the spectrum properly. Further investigations are needed to determine the status of MN112.

\section*{Acknowledgements}
We thank referee for useful comments and John Hillier for providing CMFGEN code. The reported study was funded by RFBR and NSFB according to the research project N 19-51-45001.
S.\,F. are grateful to the Russian Foundation for Basic Research (grant N 19-02-000432). A.\,K. are grateful to the Russian Foundation for Basic Research (grant N 19-02-00311). Partly based on observations obtained with the Apache Point Observatory 3.5-meter telescope, which is owned and operated by the Astrophysical Research Consortium.

\section*{Data Availability}
The data underlying this article will be shared on reasonable request to the corresponding author.




\bibliographystyle{mnras}
\bibliography{bibtexbase} 

@ARTICLE{Hillier1990,
   author = {{Hillier}, D.~J.},
    title = "{An iterative method for the solution of the statistical and radiative equilibrium equations in expanding atmospheres}",
  journal = {\aap},
     year = 1990,
    month = may,
   volume = 231,
    pages = {116-124},
   adsurl = {http://adsabs.harvard.edu/abs/1990A%26A...231..116H}
}

@ARTICLE{Hillier1998,
   author = {{Hillier}, D.~J. and {Miller}, D.~L.},
    title = "{The Treatment of Non-LTE Line Blanketing in Spherically Expanding Outflows}",
  journal = {\apj},
     year = 1998,
    month = mar,
   volume = 496,
    pages = {407-427},
      doi = {10.1086/305350},
   adsurl = {http://adsabs.harvard.edu/abs/1998ApJ...496..407H}
}

@ARTICLE{Lamers1996,
   author = {{Lamers}, H.~J.~G.~L.~M. and {Najarro}, F. and {Kudritzki}, R.~P. and 
	{Morris}, P.~W. and {Voors}, R.~H.~M. and {van Gent}, J.~I. and 
	{Waters}, L.~B.~F.~M. and {de Graauw}, T. and {Beintema}, D. and 
	{Valentijn}, E.~A. and {Hillier}, D.~J.},
    title = "{The ISO-SWS spectrum of P Cygni.}",
  journal = {\aap},
     year = 1996,
    month = nov,
   volume = 315,
    pages = {L229-L232},
   adsurl = {http://adsabs.harvard.edu/abs/1996A%26A...315L.229L}
}

@ARTICLE{Hillier1989,
   author = {{Hillier}, D.~J.},
    title = "{WC stars - Hot stars with cold winds}",
  journal = {\apj},
     year = 1989,
    month = dec,
   volume = 347,
    pages = {392-408},
      doi = {10.1086/168127},
   adsurl = {http://adsabs.harvard.edu/abs/1989ApJ...347..392H}
}

@ARTICLE{Hillier1999,
   author = {{Hillier}, D.~J. and {Miller}, D.~L.},
    title = "{Constraints on the Evolution of Massive Stars through Spectral Analysis. I. The WC5 Star HD 165763}",
  journal = {\apj},
     year = 1999,
    month = jul,
   volume = 519,
    pages = {354-371},
      doi = {10.1086/307339},
   adsurl = {http://adsabs.harvard.edu/abs/1999ApJ...519..354H}
}

@ARTICLE{Gvaramadze2010,
   author = {{Gvaramadze}, V.~V. and {Kniazev}, A.~Y. and {Fabrika}, S. and 
	{Sholukhova}, O. and {Berdnikov}, L.~N. and {Cherepashchuk}, A.~M. and 
	{Zharova}, A.~V.},
    title = "{MN112: a new Galactic candidate luminous blue variable}",
  journal = {\mnras},
archivePrefix = "arXiv",
   eprint = {0912.5080},
 primaryClass = "astro-ph.SR",
     year = 2010,
    month = jun,
   volume = 405,
    pages = {520-524},
      doi = {10.1111/j.1365-2966.2010.16469.x},
   adsurl = {http://adsabs.harvard.edu/abs/2010MNRAS.405..520G}
}

@ARTICLE{Genderen2001AA,
   author = {{van Genderen}, A.~M.},
    title = "{S Doradus variables in the Galaxy and the Magellanic Clouds}",
  journal = {\aap},
     year = 2001,
    month = feb,
   volume = 366,
    pages = {508-531},
      doi = {10.1051/0004-6361:20000022},
   adsurl = {http://adsabs.harvard.edu/abs/2001%A26A...366..508V}
}

@ARTICLE{Massey2007,
   author = {{Massey}, P. and {McNeill}, R.~T. and {Olsen}, K.~A.~G. and 
	{Hodge}, P.~W. and {Blaha}, C. and {Jacoby}, G.~H. and {Smith}, R.~C. and 
	{Strong}, S.~B.},
    title = "{A Survey of Local Group Galaxies Currently Forming Stars. III. A Search for Luminous Blue Variables and Other H{$\alpha$} Emission-Line Stars}",
  journal = {\aj},
archivePrefix = "arXiv",
   eprint = {0709.1267},
     year = 2007,
    month = dec,
   volume = 134,
    pages = {2474-2503},
      doi = {10.1086/523658},
   adsurl = {http://adsabs.harvard.edu/abs/2007AJ....134.2474M}
}

@ARTICLE{Clark2005,
   author = {{Clark}, J.~S. and {Larionov}, V.~M. and {Arkharov}, A.},
    title = "{On the population of galactic Luminous Blue Variables}",
  journal = {\aap},
     year = 2005,
    month = may,
   volume = 435,
    pages = {239-246},
      doi = {10.1051/0004-6361:20042563},
   adsurl = {http://adsabs.harvard.edu/abs/2005A%26A...435..239C}
}

@ARTICLE{Humph1994,
   author = {{Humphreys}, R.~M. and {Davidson}, K.},
    title = "{The luminous blue variables: Astrophysical geysers}",
  journal = {\pasp},
     year = 1994,
    month = oct,
   volume = 106,
    pages = {1025-1051},
      doi = {10.1086/133478},
   adsurl = {http://adsabs.harvard.edu/abs/1994PASP..106.1025H}
}

@ARTICLE{Humphreys2016,
   author = {{Humphreys}, R.~M. and {Martin}, J.~C. and {Gordon}, M.~S. and 
	{Jones}, T.~J.},
    title = "{Multiple Outflows in the Giant Eruption of a Massive Star}",
  journal = {\apj},
archivePrefix = "arXiv",
   eprint = {1606.04959},
 primaryClass = "astro-ph.SR",
     year = 2016,
    month = aug,
   volume = 826,
      eid = {191},
    pages = {191},
      doi = {10.3847/0004-637X/826/2/191},
   adsurl = {http://adsabs.harvard.edu/abs/2016ApJ...826..191H}
}

@ARTICLE{Bailer-Jones2018,
   author = {{Bailer-Jones}, C.~A.~L. and {Farnocchia}, D. and {Meech}, K.~J. and 
	{Brasser}, R. and {Micheli}, M. and {Chakrabarti}, S. and {Buie}, M.~W. and 
	{Hainaut}, O.~R.},
    title = "{Plausible Home Stars of the Interstellar Object {\lsquo}Oumuamua Found in Gaia DR2}",
  journal = {\aj},
archivePrefix = "arXiv",
   eprint = {1809.09009},
 primaryClass = "astro-ph.EP",
     year = 2018,
    month = nov,
   volume = 156,
      eid = {205},
    pages = {205},
      doi = {10.3847/1538-3881/aae3eb},
   adsurl = {http://adsabs.harvard.edu/abs/2018AJ....156..205B}
}

@ARTICLE{Najarro1997,
   author = {{Najarro}, F. and {Hillier}, D.~J. and {Stahl}, O.},
    title = "{A spectroscopic investigation of P Cygni. I. H and HeI lines.}",
  journal = {\aap},
     year = 1997,
    month = oct,
   volume = 326,
    pages = {1117-1134},
   adsurl = {http://adsabs.harvard.edu/abs/1997A%26A...326.1117N}
}

@INPROCEEDINGS{Najarro2001,
   author = {{Najarro}, F.},
    title = "{Spectroscopy of P Cygni}",
booktitle = {P Cygni 2000: 400 Years of Progress},
     year = 2001,
   series = {Astronomical Society of the Pacific Conference Series},
   volume = 233,
   editor = {{de Groot}, M. and {Sterken}, C.},
    month = jun,
    pages = {133},
   adsurl = {http://adsabs.harvard.edu/abs/2001ASPC..233..133N}
}

@ARTICLE{Balan2010,
   author = {{Balan}, A. and {Tycner}, C. and {Zavala}, R.~T. and {Benson}, J.~A. and 
	{Hutter}, D.~J. and {Templeton}, M.},
    title = "{The Spatially Resolved H{$\alpha$}-emitting Wind Structure of P Cygni}",
  journal = {\aj},
archivePrefix = "arXiv",
   eprint = {1004.0376},
 primaryClass = "astro-ph.SR",
     year = 2010,
    month = jun,
   volume = 139,
    pages = {2269-2278},
      doi = {10.1088/0004-6256/139/6/2269},
   adsurl = {http://adsabs.harvard.edu/abs/2010AJ....139.2269B}
}

@ARTICLE{LamersGroot1992,
   author = {{Lamers}, H.~J.~G.~L.~M. and {de Groot}, M.~J.~H.},
    title = "{Observed evolutionary changes in the visual magnitude of the luminous blue variable P Cygni}",
  journal = {\aap},
     year = 1992,
    month = apr,
   volume = 257,
    pages = {153-162},
   adsurl = {http://adsabs.harvard.edu/abs/1992A%26A...257..153L}
}

@ARTICLE{Stahl1991,
   author = {{Stahl}, O. and {Mandel}, H. and {Szeifert}, T. and {Wolf}, B. and 
	{Zhao}, F.},
    title = "{Forbidden emission lines in the spectrum of P Cygni}",
  journal = {\aap},
     year = 1991,
    month = apr,
   volume = 244,
    pages = {467-469},
   adsurl = {http://adsabs.harvard.edu/abs/1991A%26A...244..467S}
}

@ARTICLE{Lamers1983,
   author = {{Lamers}, H.~J.~G.~L.~M. and {de Groot}, M. and {Cassatella}, A.
	},
    title = "{The distance, temperature, and luminosity of the hypergiant P Cygni (B1 IA +)}",
  journal = {\aap},
     year = 1983,
    month = dec,
   volume = 128,
    pages = {299-310},
   adsurl = {http://adsabs.harvard.edu/abs/1983A%26A...128..299L}
}

@ARTICLE{Maryeva2018,
   author = {{Maryeva}, O. and {Koenigsberger}, G. and {Egorov}, O. and {Rossi}, C. and 
	{Polcaro}, V.~F. and {Calabresi}, M. and {Viotti}, R.~F.},
    title = "{Wind and nebula of the M 33 variable GR 290 (WR/LBV)}",
  journal = {\aap},
archivePrefix = "arXiv",
   eprint = {1804.10940},
 primaryClass = "astro-ph.SR",
     year = 2018,
    month = sep,
   volume = 617,
      eid = {A51},
    pages = {A51},
      doi = {10.1051/0004-6361/201732540},
   adsurl = {http://adsabs.harvard.edu/abs/2018A%26A...617A..51M}
}

@ARTICLE{Groh2011,
   author = {{Groh}, J.~H. and {Vink}, J.~S.},
    title = "{The bi-stability jump as the origin for multiple P-Cygni absorption components in luminous blue variables}",
  journal = {\aap},
archivePrefix = "arXiv",
   eprint = {1106.3007},
 primaryClass = "astro-ph.SR",
     year = 2011,
    month = jul,
   volume = 531,
      eid = {L10},
    pages = {L10},
      doi = {10.1051/0004-6361/201117087},
   adsurl = {http://adsabs.harvard.edu/abs/2011A%26A...531L..10G}
}

@ARTICLE{Fitzpatrick1999,
   author = {{Fitzpatrick}, E.~L.},
    title = "{Correcting for the Effects of Interstellar Extinction}",
  journal = {\pasp},
   eprint = {astro-ph/9809387},
     year = 1999,
    month = jan,
   volume = 111,
    pages = {63-75},
      doi = {10.1086/316293},
   adsurl = {http://adsabs.harvard.edu/abs/1999PASP..111...63F}
}

@ARTICLE{Smith2019,
       author = {{Smith}, Nathan and {Aghakhanloo}, Mojgan and {Murphy}, Jeremiah W. and
         {Drout}, Maria R. and {Stassun}, Keivan G. and {Groh}, Jose H.},
        title = "{On the Gaia DR2 distances for Galactic luminous blue variables}",
      journal = {\mnras},
         year = 2019,
        month = sep,
       volume = {488},
       number = {2},
        pages = {1760-1778},
          doi = {10.1093/mnras/stz1712},
archivePrefix = {arXiv},
       eprint = {1805.03298},
 primaryClass = {astro-ph.SR},
       adsurl = {https://ui.adsabs.harvard.edu/abs/2019MNRAS.488.1760S}
}

@ARTICLE{Mahy2016,
   author = {{Mahy}, L. and {Hutsem{\'e}kers}, D. and {Royer}, P. and {Waelkens}, C.
	},
    title = "{Tracing back the evolution of the candidate LBV HD 168625}",
  journal = {\aap},
archivePrefix = "arXiv",
   eprint = {1608.01087},
 primaryClass = "astro-ph.SR",
     year = 2016,
    month = oct,
   volume = 594,
      eid = {A94},
    pages = {A94},
      doi = {10.1051/0004-6361/201628584},
   adsurl = {http://adsabs.harvard.edu/abs/2016A%26A...594A..94M}
}

@ARTICLE{Stahl2001,
   author = {{Stahl}, O. and {Jankovics}, I. and {Kov{\'a}cs}, J. and {Wolf}, B. and 
	{Schmutz}, W. and {Kaufer}, A. and {Rivinius}, T. and {Szeifert}, T.
	},
    title = "{Long-term spectroscopic monitoring of the Luminous Blue Variable AG Carinae}",
  journal = {\aap},
     year = 2001,
    month = aug,
   volume = 375,
    pages = {54-69},
      doi = {10.1051/0004-6361:20010824},
   adsurl = {http://adsabs.harvard.edu/abs/2001A%26A...375...54S}
}

@ARTICLE{Hillier1991,
   author = {{Hillier}, D.~J.},
    title = "{The effects of electron scattering and wind clumping for early emission line stars}",
  journal = {\aap},
     year = 1991,
    month = jul,
   volume = 247,
    pages = {455-468},
   adsurl = {http://adsabs.harvard.edu/abs/1991A%26A...247..455H}
}

@INPROCEEDINGS{Najarro1997b,
   author = {{Najarro}, F. and {Kudritzki}, R.-P. and {Hillier}, D.~J. and 
	{Lamers}, H.~J.~G.~L.~M. and {Voors}, R.~H.~M. and {Morris}, P.~W. and 
	{Waters}, L.~B.~F.~M.},
    title = "{The ISO-SWS Spectrum of P Cygni}",
booktitle = {Luminous Blue Variables: Massive Stars in Transition},
     year = 1997,
   series = {Astronomical Society of the Pacific Conference Series},
   volume = 120,
   editor = {{Nota}, A. and {Lamers}, H.},
    pages = {105},
   adsurl = {http://adsabs.harvard.edu/abs/1997ASPC..120..105N}
}

@ARTICLE{Groh2009b,
   author = {{Groh}, J.~H. and {Damineli}, A. and {Hillier}, D.~J. and {Barb{\'a}}, R. and 
	{Fern{\'a}ndez-Laj{\'u}s}, E. and {Gamen}, R.~C. and {Mois{\'e}s}, A.~P. and 
	{Solivella}, G. and {Teodoro}, M.},
    title = "{Bona Fide, Strong-Variable Galactic Luminous Blue Variable Stars are Fast Rotators: Detection of a High Rotational Velocity in HR Carinae}",
  journal = {\apjl},
archivePrefix = "arXiv",
   eprint = {0909.4459},
 primaryClass = "astro-ph.SR",
     year = 2009,
    month = nov,
   volume = 705,
    pages = {L25-L30},
      doi = {10.1088/0004-637X/705/1/L25},
   adsurl = {http://adsabs.harvard.edu/abs/2009ApJ...705L..25G}
}

@ARTICLE{Humphreys2017,
   author = {{Humphreys}, R.~M. and {Gordon}, M.~S. and {Martin}, J.~C. and 
	{Weis}, K. and {Hahn}, D.},
    title = "{Luminous and Variable Stars in M31 and M33. IV. Luminous Blue Variables, Candidate LBVs, B[e] Supergiants, and the Warm Hypergiants: How to Tell Them Apart}",
  journal = {\apj},
archivePrefix = "arXiv",
   eprint = {1611.07986},
 primaryClass = "astro-ph.SR",
     year = 2017,
    month = feb,
   volume = 836,
      eid = {64},
    pages = {64},
      doi = {10.3847/1538-4357/aa582e},
   adsurl = {https://ui.adsabs.harvard.edu/abs/2017ApJ...836...64H}
}

@ARTICLE{Mehner2017,
       author = {{Mehner}, A. and {Baade}, D. and {Groh}, J.~H. and {Rivinius}, T. and
         {Hambsch}, F. -J. and {Bartlett}, E.~S. and {Asmus}, D. and
         {Agliozzo}, C. and {Szeifert}, T. and {Stahl}, O.},
        title = "{Spectroscopic and photometric oscillatory envelope variability during the S Doradus outburst of the luminous blue variable R71}",
      journal = {\aap},
         year = "2017",
        month = "Dec",
       volume = {608},
          eid = {A124},
        pages = {A124},
          doi = {10.1051/0004-6361/201731829},
archivePrefix = {arXiv},
       eprint = {1709.00160},
 primaryClass = {astro-ph.SR},
       adsurl = {https://ui.adsabs.harvard.edu/abs/2017A&A...608A.124M}
}

@INPROCEEDINGS{Wilson2004,
author = {{Wilson}, J.~C. and {Henderson}, C.~P. and {Herter}, T.~L. and
{Matthews}, K. and {Skrutskie}, M.~F. and {Adams}, J.~D. and
{Moon}, D.-S. and {Smith}, R. and {Gautier}, N. and {Ressler}, M. and
{Soifer}, B.~T. and {Lin}, S. and {Howard}, J. and {LaMarr}, J. and
{Stolberg}, T.~M. and {Zink}, J.},
title = "{Mass producing an efficient NIR spectrograph}",
booktitle = {Ground-based Instrumentation for Astronomy},
year = 2004,
series = {\procspie},
volume = 5492,
editor = {{Moorwood}, A.~F.~M. and {Iye}, M.},
month = sep,
pages = {1295-1305},
doi = {10.1117/12.550925},
adsurl = {http://adsabs.harvard.edu/abs/2004SPIE.5492.1295W}
}

@INPROCEEDINGS{Owocki1991,
       author = {{Owocki}, S.~P.},
        title = "{A Smooth Source Function Method for Including Scattering in Radiatively Driven Wind Simulations}",
    booktitle = {NATO Advanced Science Institutes (ASI) Series C},
         year = 1991,
       editor = {{Crivellari}, Lucio and {Hubeny}, I. and {Hummer}, D.~G.},
       series = {NATO Advanced Science Institutes (ASI) Series C},
       volume = {341},
        month = jan,
        pages = {235},
       adsurl = {https://ui.adsabs.harvard.edu/abs/1991ASIC..341..235O}
}

@ARTICLE{Berlanas2019,
       author = {{Berlanas}, S.~R. and {Wright}, N.~J. and {Herrero}, A. and
         {Drew}, J.~E. and {Lennon}, D.~J.},
        title = "{Disentangling the spatial substructure of Cygnus OB2 from Gaia DR2}",
      journal = {\mnras},
         year = 2019,
        month = apr,
       volume = {484},
       number = {2},
        pages = {1838-1842},
          doi = {10.1093/mnras/stz117},
archivePrefix = {arXiv},
       eprint = {1901.02959},
 primaryClass = {astro-ph.SR},
       adsurl = {https://ui.adsabs.harvard.edu/abs/2019MNRAS.484.1838B}
}

@ARTICLE{Billot2010,
       author = {{Billot}, N. and {Noriega-Crespo}, A. and {Carey}, S. and {Guieu}, S. and
         {Shenoy}, S. and {Paladini}, R. and {Latter}, W.},
        title = "{Young Stellar Objects and Triggered Star Formation in The Vulpecula OB Association}",
      journal = {\apj},
         year = 2010,
        month = apr,
       volume = {712},
       number = {2},
        pages = {797-812},
          doi = {10.1088/0004-637X/712/2/797},
archivePrefix = {arXiv},
       eprint = {1003.0866},
 primaryClass = {astro-ph.GA},
       adsurl = {https://ui.adsabs.harvard.edu/abs/2010ApJ...712..797B}
}

@ARTICLE{Kharchenko2005,
       author = {{Kharchenko}, N.~V. and {Piskunov}, A.~E. and {R{\"o}ser}, S. and
         {Schilbach}, E. and {Scholz}, R. -D.},
        title = "{Astrophysical parameters of Galactic open clusters}",
      journal = {\aap},
         year = 2005,
        month = aug,
       volume = {438},
       number = {3},
        pages = {1163-1173},
          doi = {10.1051/0004-6361:20042523},
archivePrefix = {arXiv},
       eprint = {astro-ph/0501674},
 primaryClass = {astro-ph},
       adsurl = {https://ui.adsabs.harvard.edu/abs/2005A&A...438.1163K}
}

@ARTICLE{Sagar1981,
       author = {{Sagar}, R. and {Joshi}, U.~C.},
        title = "{Study of the galactic cluster NGC 6823}",
      journal = {\apss},
         year = 1981,
        month = apr,
       volume = {75},
       number = {2},
        pages = {465-472},
          doi = {10.1007/BF00648656},
       adsurl = {https://ui.adsabs.harvard.edu/abs/1981Ap&SS..75..465S}
}

@ARTICLE{Humphreys2014,
       author = {{Humphreys}, Roberta M. and {Weis}, Kerstin and {Davidson}, Kris and
         {Bomans}, D.~J. and {Burggraf}, Birgitta},
        title = "{Luminous and Variable Stars in M31 and M33. II. Luminous Blue Variables, Candidate LBVs, Fe II Emission Line Stars, and Other Supergiants}",
      journal = {\apj},
         year = 2014,
        month = jul,
       volume = {790},
       number = {1},
          eid = {48},
        pages = {48},
          doi = {10.1088/0004-637X/790/1/48},
archivePrefix = {arXiv},
       eprint = {1407.2259},
 primaryClass = {astro-ph.SR},
       adsurl = {https://ui.adsabs.harvard.edu/abs/2014ApJ...790...48H}
}




\bsp	
\label{lastpage}
\end{document}